# Visualizing Millisecond Atomic Dynamics of Nanocrystals in Liquid

Sungsu Kang[1,2], Jinho Rhee[3], Joodeok Kim[3], Sam Oaks-Leaf[4], Minwoo Kim[3], Shengsong Yang[1,2], Chang Liu[1,2], Dongsu Kim[3], Sungin Kim[3], Binyu Wu[5], Won Bo Lee[3], David T. Limmer[4,6,7,8], A. Paul Alivisatos[1,2,5]*, Peter Ercius[9]*, and Jungwon Park[3,10,11,12,13]*

[1]Department of Chemistry, University of Chicago; Chicago, Illinois 60637, United States.

[2]James Franck Institute, University of Chicago; Chicago, Illinois 60637, United States.

[3]Department of Chemical and Biological Engineering, and Institute of Chemical Processes; Seoul National University, Seoul 08826, Republic of Korea.

[4]Department of Chemistry, University of California; Berkeley, California 94720, United States.

[5]Pritzker School of Molecular Engineering, University of Chicago; Chicago, Illinois 60637, United States.

[6]Materials Sciences Division, Lawrence Berkeley National Laboratory; Berkeley, California 94720, United States.

[7]Kavli Energy NanoScience Institute, Berkeley; California 94720, United States

[8]Chemical Sciences Division, Lawrence Berkeley National Laboratory; Berkeley, California 94720, United States.

[9]Molecular Foundry, Lawrence Berkeley National Laboratory; Berkeley, California 94720, United States.

[10]Center for Nanoparticle Research, Institute for Basic Science (IBS); Seoul 08826, Republic of Korea.

[11]Institute of Engineering Research, College of Engineering; Seoul National University, Seoul 08826, Republic of Korea.

[12]Advanced Institute of Convergence Technology, Seoul National University; Suwon 16229, Republic of Korea.

[13]Hyundai Motor Group-Seoul National University (HMG-SNU) Joint Battery Research Center (JBRC), Seoul National University; Seoul 08826, Republic of Korea.

*Corresponding authors. Email: jungwonpark@snu.ac.kr (J.P.); percuis@lbl.gov (P.E.); paul.alivisatos@uchicago.edu (A.P.A).

## Abstract

Atomic structures of nanomaterials are inherently dynamic, continuously reshaped through interactions with chemical species and external stimuli. Such dynamics are further amplified as the size and dimensionality of nanomaterials are reduced. Despite advances in analytical methods, it remains challenging to capture structural dynamics of nanomaterials in reactive environments with both atomic spatial resolution and commensurate temporal resolution. Here, we directly visualize atomic-scale dynamics of gold (Au) nanocrystals in reactive liquid environments with millisecond-speed liquid cell electron microscopy (EM) and deep-learning denoising. We uncover reversible fluctuations in local crystallinity of Au nanocrystals dependent on the surrounding chemical environment. These transient fluctuations, driven by interactions at nanocrystal-liquid interfaces, critically influence dissolution kinetics and grain boundary relaxation. By overcoming the spatiotemporal limitations in conventional liquid cell EM, our findings provide insights into how transient nanoscale structures dictate the stability and reactivity of nanomaterials.

## Main

Atomic structures of nanocrystals govern their catalytic activity[1], electronic properties[2], and stability[3]. Advanced spectroscopic[4–6] and microscopic[7,8] techniques have elucidated static atomic structures of nanocrystals. Yet, throughout their formation and utilization, nanocrystals constantly interact with solvents, ligands, and chemical reactants, leading to dynamic changes in atomic arrangements. Because surface atoms have low coordination and are exposed to the surrounding chemical species, they are particularly labile and prone to initiating such changes[9–12]. These processes can be further accelerated under external stimuli, including heat[13], electrical bias[14], or (electro)chemical potential[15]. Thus, the intrinsic dynamical nature of nanocrystal structures calls for advanced analytical methods capable of tracking atomic configurations under realistic chemical environments on relevant time scales.

Structural transformations of nanometer-sized crystals are expected to transform on the millisecond or faster time scales[16]. However, capturing such atomic-scale dynamics on these time scales remains a challenge for both theoretical and experimental approaches. Even in molecular dynamics (MD) simulations, accurate models for nanometer-sized materials have only recently emerged[17,18], and representing the full chemistry of nanocrystals in solution is still computationally demanding. Recent advances in electron microscopy (EM) have enabled imaging single nanocrystals in solution at high-resolution by encapsulating liquid samples between ultrathin films such as graphene, thereby preserving the solution state in high vacuum during *in situ* imaging[19–21]. Liquid cell EM has revealed processes such as nanocrystal growth, etching, and ripening in time-resolved image series[22–26]. However, the structural information extracted from individual images is averaged over relatively long electron exposure times, which typically exceed the timescale of the reactions. Furthermore, EM of liquid samples suffers from strong background contrast originating from the solvent and window materials[27]. These challenges persist even in state-of-the-art liquid cell EM with aberration-corrected lenses, atomically thin window materials, and direct electron detectors. This points to the compelling need to further improve liquid cell EM techniques to simultaneously achieve atomic spatial resolution and millisecond or faster temporal resolution.

Here, we directly visualize fast atomic-scale dynamics of gold (Au) nanocrystals in a reactive solution environment through millisecond-speed atomic-resolution liquid cell EM. We reveal that surface coordination, exposed to chemical environment, induces millisecond-scale structural fluctuations between crystalline and disordered local atomic arrangements, serving as a new pathway in nanocrystal transformations, such as etching and grain boundary relaxation. By simultaneously achieving millisecond temporal and atomic spatial resolution, our work uncovers structural fluctuations in nanocrystals that exist but have not been approached in studies based on conventional analytical techniques.

**Millisecond-speed atomic-resolution liquid cell EM of nanocrystals**

Achieving high spatiotemporal resolution in liquid cell EM is hindered by fundamental trade-offs in imaging principles[28]. Electron dose rates required for high-resolution imaging of inorganic materials, typically $10^3$–$10^4$ e$^-$ Å$^{-2}$ s$^{-1}$, yield low electron doses of 1–10 e$^-$ Å$^{-2}$ per frame at millisecond frame rates. Under this condition, the images suffer from low signal-to-noise ratio (SNR), which is further reduced in liquid cell EM due to electron scattering from liquid and window materials even in graphene liquid cells (GLCs). To overcome these limitations, we applied a neural-network-based image processing procedure (Extended Data Fig. 1a). We first denoised a raw dataset using a 3×3 blind-spot neural network to remove pixel-correlated noises typical of EM detectors[29]. The neural-network-denoising reveals that a GLC-encapsulated solution generates an amorphous background that interferes with atomic-scale nanocrystal structures. A Butterworth band-reject filter was applied to selectively suppress the GLC background, allowing for clear visualization of atomic-scale nanocrystal structures while maintaining the millisecond temporal resolution from data acquisition.

## Atomic-scale structural fluctuations of Au nanocrystals in etching solution

Using millisecond-speed atomic-resolution GLC EM, we study the structural dynamics of small Au nanocrystals etched in aqueous $FeCl_3$ solutions (Methods). Under electron beam irradiation, etching proceeds through the complexation of surface Au with $Cl^-$, mediated by a $Fe^{2+}/Fe^{3+}$ redox couple, to form soluble $[AuCl_2]^-$ (Fig. 1a, Supplementary Note 1)[30–33]. Understanding complexation-mediated etching is important, as it is a primary mechanism in many wet chemical etching processes[34,35]. Further, while previous liquid cell EM studies utilized etching to probe nanocrystal properties such as kinetic stabilization, relative surface energies, and chemical reactivity[30–33,36,37], how nanocrystal surfaces interact with the solvent has remained elusive.

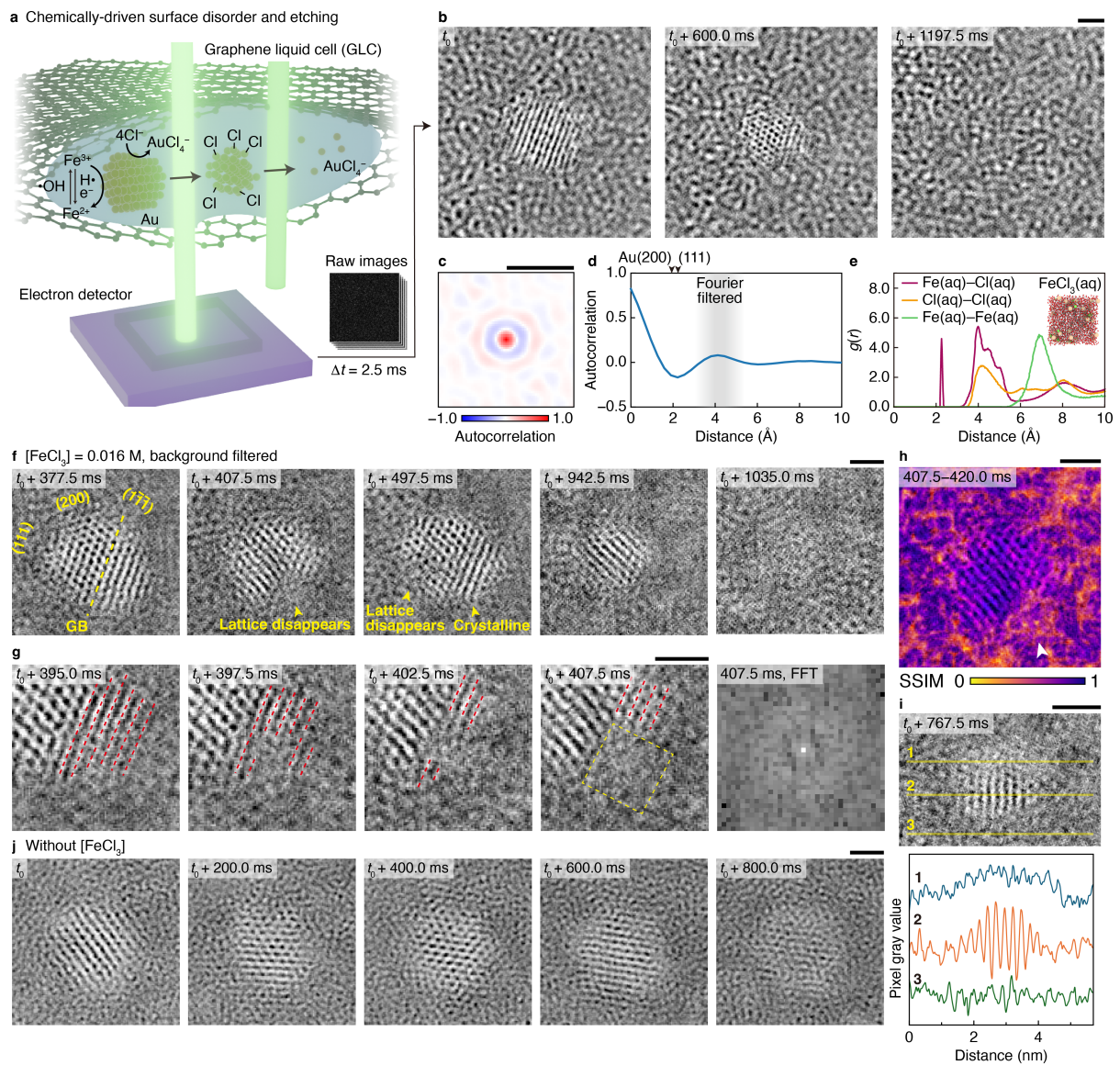

**Fig. 1 | Imaging fast atomic-scale fluctuation in Au nanocrystals interacting with solution environment. a**, Schematic of Au nanocrystals interacting with a $FeCl_3$(aq) solution encapsulated in a GLC. Au nanocrystals are chemically etched through a redox reaction

mediated with a $Fe^{3+}/Fe^{2+}$ ion pair and a complexation reaction with $Cl^-$, initiated by electron beam irradiation. Raw TEM image series, captured at 2.5 ms temporal resolution, suffer from high levels of noise, which are subsequently removed by a self-supervised denoising framework. **b**, Representative snapshots from the denoised TEM image series showing complete dissolution of a ~3 nm Au nanocrystal within ~1.2 s in a [$FeCl_3$(aq)] = 0.016 M solution under an electron dose rate of 1,500 $e^-/Å^2$ s. **c,d**, Autocorrelated image of a background region after a complete dissolution at $t_0$ + 1197.5 ms (c), and corresponding radial average (d). The background signals near the first maximum are FFT filtered. **e**, Radial distribution function $g(r)$ between different elements in simulated $FeCl_3$(aq) solution. **f**, Background-filtered TEM image series, showing disappearance and reappearance of Au nanocrystal lattice during etching. GB: grain boundary. **g**, Millisecond-scale images showing lattice fluctuation in a magnified view, and a FFT of the region marked with a yellow box where lattice transiently disappears. (111) planes are marked with red lines. **h**, SSIM of the images over 407.5−420.0 ms, overlayed with the frame at 407 ms. **i**, A background-filtered image frame (upper panel) showing nanocrystals exhibiting structural fluctuation, and pixel intensity profiles (lower panels) along the lines labeled 1, 2, and 3. **j**, Denoised- and background-filtered TEM image series showing an Au nanocrystal in a solution without $FeCl_3$. Scale bars, 10 Å.

We prepared GLCs encapsulating an aqueous buffer solution containing Au nanocrystals and $FeCl_3$. The edges of liquid pockets where the liquid is thinnest were imaged to obtain highest possible spatial resolution (Extended Data Fig. 2a–c). The concentration of radiolytic species and the temperature remain nearly constant in the time scale of our observation (Extended Data Fig. 3, and Supplementary Note 1 and 2). Our image processing procedure recovers clear lattice structures of dissolving Au nanocrystals from TEM image series, captured at 2.5 ms/frame (Fig. 1b, Methods, Extended Data Fig. 1a). Mean SNR improves from –14±2 to 4±2 dB in the raw and denoised image series, respectively (Extended Data Fig. 1c–f and Methods)[38]. The GLC background, broadly correlated around ~4 Å, was also suppressed (Fig. 1c,d, and Extended Data Fig. 1b). MD simulations indicate that this background likely arises from solvated Fe and Cl ions, showing radial distribution function peaks near ~4 Å (Fig. 1e).

This optimized workflow enabled the direct visualization of atomic structural changes in a ~3 nm Au nanocrystal (Fig. 1f and Supplementary Movie 1). Under our experimental conditions, Au nanocrystals of similar size are typically fully etched within ~1 s. Slow rotation of nanocrystals in liquid cells enables direct comparison between image frames (Extended Data Fig. 2d–g and Supplementary Note 3). We observed intermittent disappearance and reappearance of lattice fringes near the nanocrystal surface occurring on millisecond time scales during the etching process. For example, the nanocrystal which is fully crystalline at 377.5 ms from the start of the observation ($t_0$), loses lattice contrast in the lower-right region by 407.5 ms (Fig. 1f). The lattice contrast subsequently recovers by 497.5 ms. The left region also loses contrast at 497.5 ms and recovers by 537.5 ms. Structural similarity index measure (SSIM), which quantifies similarity between two images, exhibits lower values in this region compared to the crystalline lattice (Fig. 1h). Comparison of the averaged, denoised, and filtered images indicates that regions without lattice contrast are not artifacts from image processing (Extended Data Fig. 4a). Pixel intensity analysis further confirms the presence of mass in these regions evidenced by extra scattering compared to adjacent liquid regions (Fig. 1i), pointing to transient and dynamic atomic fluctuation between crystalline and disordered atomic arrangements[10].

Au nanocrystals in solutions without $FeCl_3$ do not exhibit such fluctuations (Fig. 1j), indicating that the chemical environment promotes the fluctuation under electron beam. Strong coordination of $Cl^-$ on the Au surface, revealed by a previous experiment[39], can weaken the underlying Au−Au bonds, making the surface Au atoms easier to deviate from lattice positions. Previous *in situ* TEM observations also indicated that the adsorption of gaseous molecules to small Pt nanocrystals induce fluxional behavior[40]. Thus, the chemical coordination may decrease the energy barrier between the crystalline and disordered configurations and change their relative energies, while the electron beam provides driving force for the transitions between the two configurations. In our solution etching system, we can further establish relationships between the local chemical environment and structural dynamics by analyzing nanocrystal etching trajectories.

**Quantifying etching trajectories and structural fluctuations**

To quantify etching trajectories, we tracked the crystalline diameter of individual nanocrystals over time. The crystalline diameter of each frame is measured by masking the strongest FFT peak in the fast Fourier transform (FFT) and applying an inverse FFT to map the crystalline domain[10]. We excluded frames where crystalline peaks were not detected or where only partial domains of polycrystalline nanocrystals were resolved. The crystalline diameter decreases increasingly faster over time (Fig. 2a, blue circles), consistent with the Kelvin equation: $dr/dt = kS_b/\rho \exp(r_c/r)$, where $r$, $t$, $k$, $S_b$, $\rho$, and $r_c$ denote a particle radius, time, constant related to the dissolution environment, bulk solubility, density, and critical radius, respectively[41,42]. The critical radius, $r_c$, is proportional to the surface energy, $\gamma$, by $r_c = 2\gamma V_m/RT$.

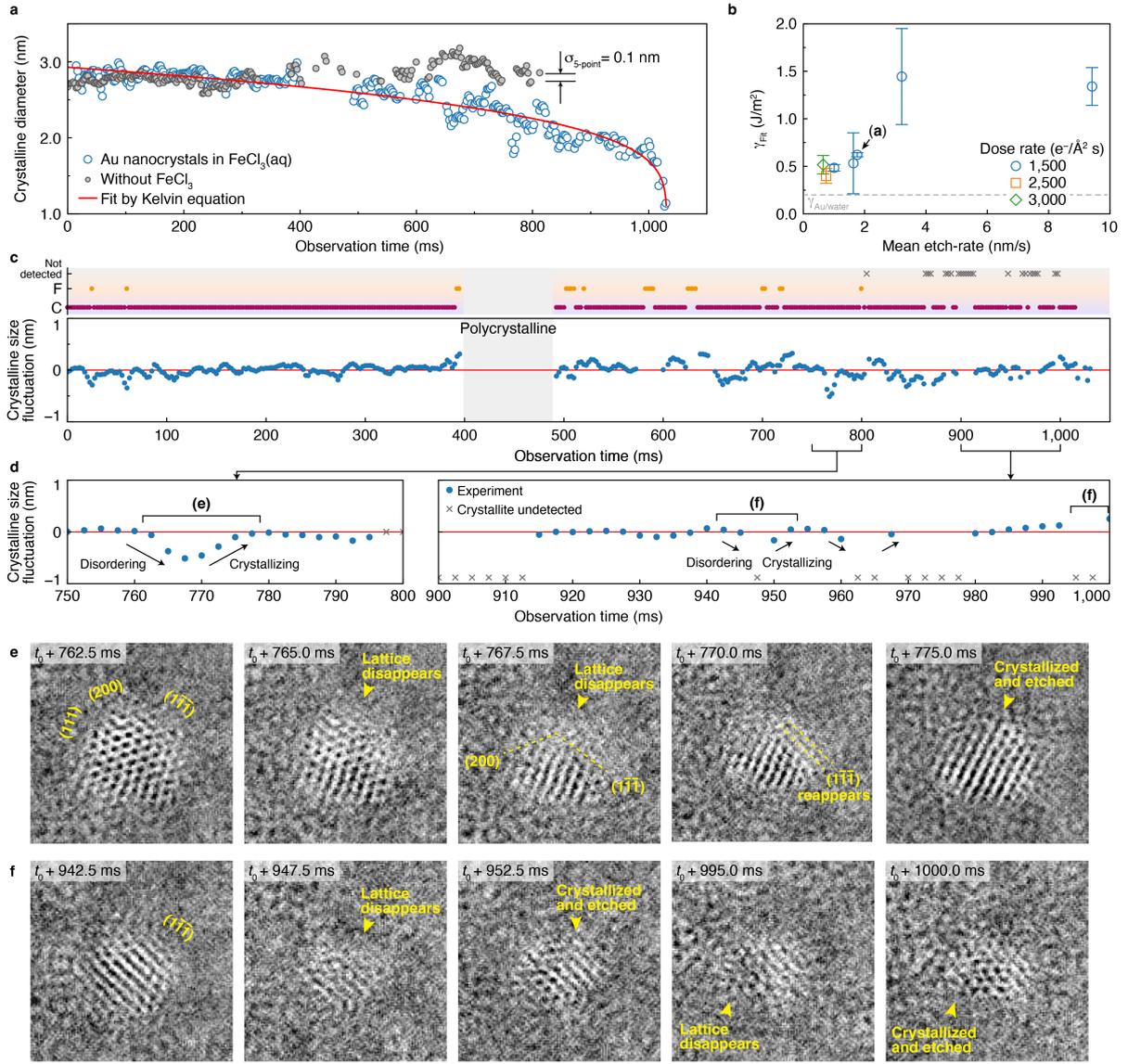

**Fig. 2 | Structural fluctuation between crystalline and disordered states. a**, Crystalline diameter of nanocrystals over time. (blue circles) Experimental data measured from the Au nanocrystal dissolving in the $FeCl_3$ solution, shown in Fig. 1f. (red line) Fit using the Kelvin equation where the effective interfacial energy $\gamma_{Fit}$ is fitted to be 0.271±0.005 J/m². (gray symbols) Experimental data measured from the Au nanocrystal in the solution without $FeCl_3$ in Fig. 1j. 5-frame standard deviation of the measured crystalline diameters are smaller than 0.1 nm (Extended Data Fig. 4). **b**, Relationship between $\gamma_{Fit}$ and mean etch-rates of different Au nanocrystals imaged under different electron dose rates. $\gamma_{Au/water}$ = ~0.2 J/m² is denoted as a dashed line. Error bars denote one-standard-deviation errors in the fitted parameters. **c**, Quantitative analysis of structural fluctuation. (upper panel) Identification of frames when a nanocrystal is fully crystalline (C) and exhibits crystallinity fluctuation (F). Frames without detectable FFT peaks are also indicated. (lower panel) Deviation of the crystalline diameter from the fit predicted by the Kelvin equation over the observation. **d**, Expanded views of selected time intervals (750–800 and 900–1000 ms), showing intermittent drops in crystalline diameter. Diameters are not measured during periods when crystalline FFT peaks are

undetectable (× symbols). **e,f,** TEM snapshots corresponding to the fluctuation periods highlighted in (d), showing transitions between crystalline and disordered regions. Scale bars, 10 Å.

The red curve in Figure 2a shows a fit of the etching trajectory of the nanocrystal shown in Fig. 1. This provides an effective nanocrystal-solution interfacial energy $\gamma_{Fit}$ = ~0.271±0.005 J/m$^2$. The value is slightly higher than Au/water interfacial energy, $\gamma_{Au/water}$ = ~0.2 J/m$^2$ calculated by density functional theory (DFT)[43], reflecting the reactive FeCl$_3$ environment. Tracking etching trajectories of seven more Au nanocrystal with initial diameters of ~ 3 nm confirms a positive correlation between $\gamma_{Fit}$ and the mean etch-rate of each nanocrystal (Fig. 2b, Extended Data Figs. 5 and 6). This correlation is likely because both $\gamma_{Fit}$ and the etch-rate depend on the local concentrations of the complexation agent, Cl$^-$. Interestingly, we observed that etch-rates vary even among nanocrystals within the same liquid pocket, which implies local chemical heterogeneity[44]. Thus, despite the initial FeCl$_3$ concentration used to prepare solutions, local Cl$^-$ concentration near different nanocrystal can vary, and the etch-rate and $\gamma_{Fit}$ are more accurate in describing this local chemical environment[45].

When part of a nanocrystal becomes disordered, the total crystalline area fluctuates and decreases accordingly. The crystalline diameter plot of the Au nanocrystal in the FeCl$_3$ solution shows larger, short-time (10−20 ms) fluctuations than in the solution without FeCl$_3$ (Fig. 2a). 5-frame standard deviations of the crystalline diameter in the non-etching solution is smaller than 0.1 nm (Extended Data Fig. 4). Thus, we identified frames with structural fluctuations where the 5-frame standard deviation of the crystalline diameter exceeds 0.1 nm. The fluctuations happen more frequently as etching proceeds (Fig. 2c, upper panel). We further quantified the fluctuations using deviations of the crystalline diameter from the Kelvin equation, which show intermittent drops and transient decreases in the crystalline area particularly in the later stages of etching (Fig. 2c, lower panel). For example, we observed transient drops in crystalline size lasting ~10–15 ms, followed by rapid recovery (Fig. 2d, left panel). Corresponding TEM images show that a fluctuation initiates at the edge where exposed (200) and (1$\bar{1}\bar{1}$) surfaces meet at 765.0 ms (Fig. 2e). By 767.5 ms, the fluctuation propagates into the upper-right ~1 nm$^2$ region, bounded by (1$\bar{1}\bar{1}$) and (200) lattice planes (dashed lines). The disordered region recrystallizes into the FCC structure through reorganization of (1$\bar{1}\bar{1}$) planes from crystalline-disordered interface by 775.0 ms. These structural fluctuations become more frequent as etching proceeds, indicated from repeated decreases and increases in crystalline diameters over 942.5–952.5 ms and 957.5–967.5 ms (Fig. 2d, right panel; Fig. 2f). Frames where the nanocrystal is mostly disordered are marked with × symbols in Fig. 2d. The rates of disordering and recrystallization are comparable, with both transitions occurring over 7.5–10 ms. Previous imaging techniques required summation of images reducing framerate such that these types of dynamics were not detectable.

**Solution environment-driven atomic structural fluctuations**

We hypothesized that the structural fluctuations occur more frequently in solutions with higher $Cl^-$ concentrations, which facilitate etching by complexation[32]. We calculated the temporal fraction of frames exhibiting structural fluctuations, $f_{fluctuation}$, over 25 ms and plotted it against the instantaneous etch-rate derived from the Kelvin equation (Fig. 3a). While the fluctuations occur across various etch-rates, the tendency to stay in the crystalline state diminishes at higher rates. Histograms of $f_{fluctuation}$, over etch-rate ranges of 0–1, 1–2, 2–3, and 3–4 nm/s, further demonstrate that structural fluctuations become more prevalent for nanocrystals etched at faster rates (Fig. 3b). $f_{fluctuation}$ over the entire observation period for each nanocrystal is correlated with the fitted interfacial energies, $\gamma_{Fit}$ (Fig. 3c). Extrapolating the linear fit of total $f_{fluctuation}$ versus $\gamma_{Fit}$ to the zero-fluctuation limit yields $\gamma = \sim 0.2$ J/m$^2$, a value consistent with the calculated Au/water interfacial energy.

Comparable rates for local disordering and recrystallization (Fig. 2d) indicate that the energy barrier for fluctuations is comparably lower than the driving force, and allows for describing relative energetics between the two configurations (Fig. 3b, insets). By calculating the ratio of frames exhibiting fluctuations and a fully crystalline state, $N_{fluctuation}/N_{crystalline}$ (Fig. 3d), the relative stability of these states can be mapped (Fig. 3e). In a fast-etching regime (3–4 nm/s), undergoing structural fluctuations are favored over maintaining crystallinity, consistent with the disappearance of fully crystalline particles [(iv) in Fig. 3b,d, and e]. The diameters of the nanocrystals in this regime are around ~1–1.5 nm. While previous experimental and theoretical studies predicted that sub-1-nm metal particles adopt disordered configurations with reduced surface energy, our observations on atomic-scale etching trajectories further highlights the importance of chemical environments on the transitions[46].

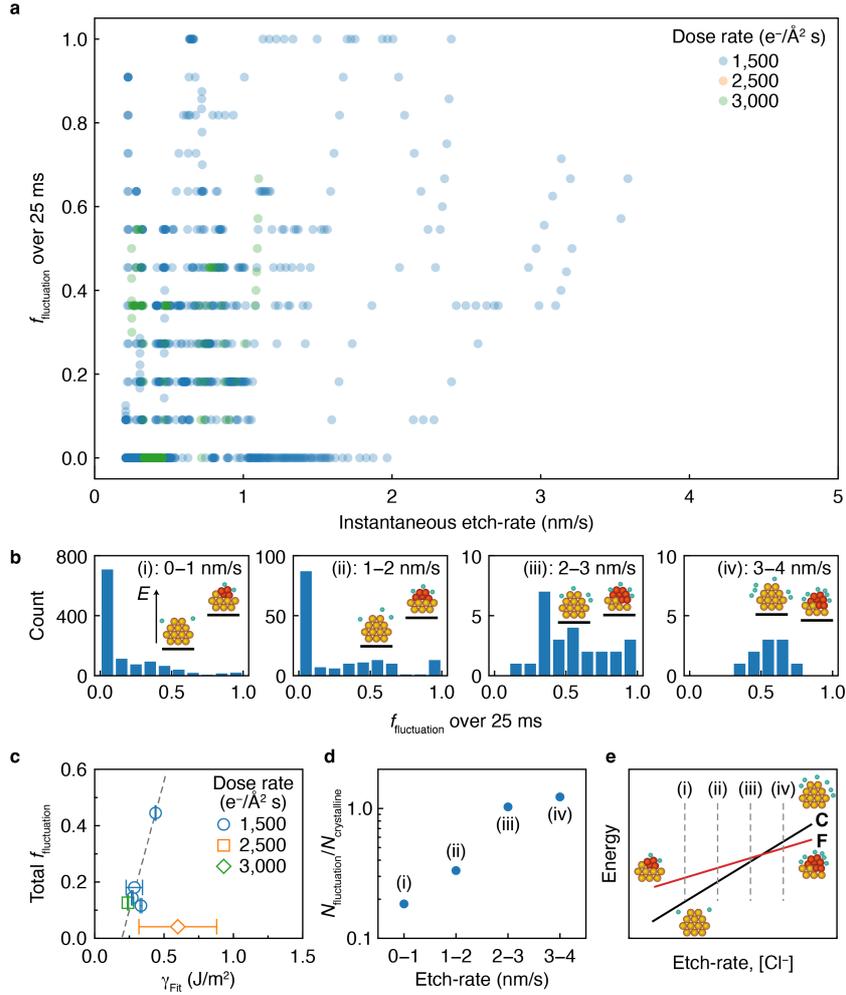

**Fig. 3 | Correlation between etching trajectories and structural fluctuations. a**, Relationship between fraction of frames exhibiting structural fluctuations, $f_{fluctuation}$, and the instantaneous etch-rate. The etch rates were derived from a Kelvin equation fit. $f_{fluctuation}$ is calculated over 25.0 ms. **b**, Histograms of $f_{fluctuation}$ calculated for the etch-rate ranges of 0–1, 1–2, 2–3, and 3–4 nm/s. Insets schematically illustrate the relative stability between crystalline and disordered states inferred from the observed populations. **c**, Correlation between the $f_{fluctuation}$ calculated during entire observation periods and the $\gamma_{Fit}$ of individual nanocrystals. (dashed line) Linear fit. **d**, Ratio of frames showing structural fluctuations to fully crystalline nanocrystal, $N_{fluctuation}/N_{crystalline}$, as a function of etch-rate, corresponding to the four ranges (i–iv) in (b). **e**, Proposed relative energetic stability between crystalline (C) and fluctuating (F) states depending on etch-rate or $Cl^-$ concentration.

## Structural fluctuation induced etching pathway

These structural fluctuations expose disordered nanocrystal surfaces where undercoordinated atoms are more susceptible to desorption[47]. We found the preferential etching of fluctuating regions from 762.5 to 772.5 ms, accompanied by changes in the crystalline diameter from ~2.3 to ~1.8 and back to ~2.0 nm (Fig. 2c). These structural fluctuations, which occur throughout our observations, accelerate etching. While conventional atom-by-atom removal, a well-known mechanism for surface etching, still occurs (Extended Data Fig. 7 and Supplementary Note 4), our observation indicates that structural fluctuations enhanced by a local chemical environment constitute a distinct and high-rate etching pathway.

**Structural fluctuation-mediated grain boundary relaxation in nanocrystals**

We further demonstrate that structural fluctuations in a polycrystalline nanocrystal provides a pathway for grain boundary relaxation. We tracked the evolution of nanocrystal structure with an asymmetrically-located (111) grain boundary, marked with a vertical dashed line at 392.5 ms in Fig. 4. The nanocrystal consists of a major domain showing [110] zone axis and a misaligned smaller domain showing only (111) lattice planes.

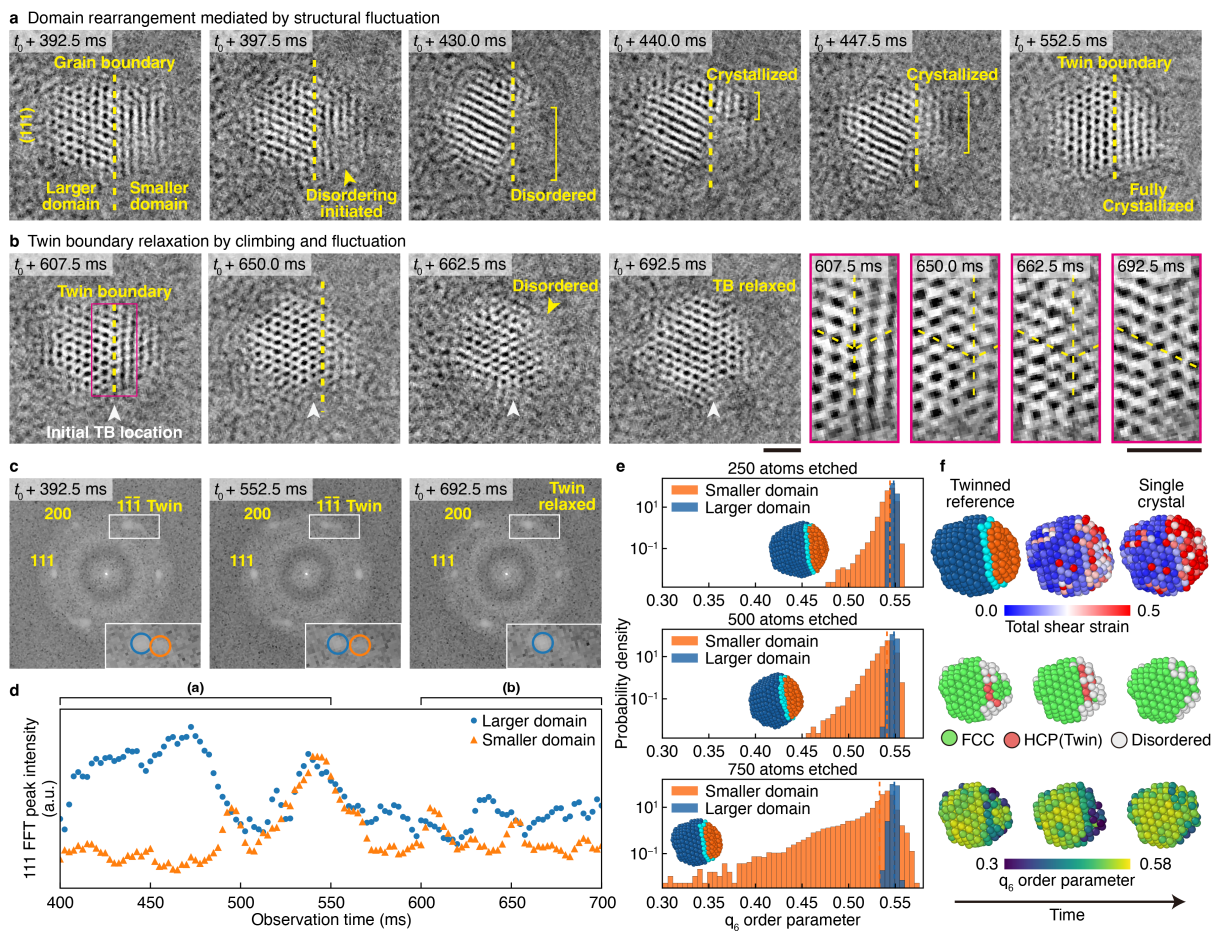

**Fig. 4 | Grain boundary relaxation mediated by structural fluctuation. a**, Millisecond-resolved TEM snapshots showing structural fluctuation of a smaller grain domain, leading to the formation of a twin boundary. **b**, Subsequent relaxation of the twin boundary via lattice-plane climbing and further fluctuation. Magnified views of the red-boxed region around the initial twin boundary are shown on the right. **c**, FFTs of selected frames reveal distinct $1\bar{1}\bar{1}$ peaks for the twin domains at 392.5 and 552.5 ms, and a single peak for the fully relaxed single-crystalline structure at 692.5 ms. Insets show magnified views of the FFT region, highlighting

peak splitting from the larger (blue) and smaller (orange) domains. **d**, Time evolution of the FFT peak intensities for the larger and smaller domains, showing domain-specific fluctuation behavior. **e**, Distributions of the average $q_6$ order parameter within smaller (orange) and large (blue) domains during MD trajectories after removing 250 (upper panel), 500 (middle panel), and 750 atoms (lower panel) from a twinned Au nanocrystal by kMC. Inset images are examples of nanocrystal models at the three stages, with atoms in smaller domains (orange), larger domains (dark blue), and at twin boundaries (cyan). **f**, Evolution of total shear strain (upper panel), crystal structure (middle panel), and $q_6$ order parameter (lower panel) of each atom within a twinned nanocrystal transforming to a single crystal after 750 atoms are removed. Atoms with fewer than 12 nearest neighbors are removed in the middle and bottom panels. Scale bars, 10 Å.

This grain boundary relaxes through a two-step process mediated by the fluctuation of the minor domain. First, the asymmetric grain boundary rearranges into a twin structure. The minor domain becomes disordered from the lower part (Fig. 4a, 397.5–440.0 ms), followed by recrystallization from the upper part (Fig. 4a, 440.0–552.5 ms). At 552.5 ms, the nanocrystal adopts a twin structure, confirmed by two distinct [1$\bar{1}\bar{1}$] FFT peaks (Fig. 4a,c). During the reconfiguration of the grain boundary to twin boundary from ~400 to ~480 ms, the disordered state persists in the minor domain as shown by weak [111] reflection intensities (Fig. 4d). In the second step, the twin structure relaxes into a single crystal. The twin boundary first migrates toward a surface by one (111) lattice plane (Fig. 4b, 607.5–650.0 ms). Then, the smaller domain becomes disordered and subsequently recrystallized (Fig. 4b, 662.5–692.5 ms), eventually adopting a single-crystalline structure.

We employed kinetic Monte Carlo (kMC) coupled with MD simulations using a neural network potential to identify the driving force for fluctuation-mediated relaxation. We modeled the etching of a ~3 nm Au nanocrystal with an asymmetrically-located (111) twin boundary (Methods, Extended Data Figs. S8–10, and Supplementary Note 5). While kMC produces locally-relaxed metastable intermediate structures during etching, MD allows for assessing disordering of each crystalline domain initialized from configurations produced by kMC through collective motion. Quantifying local order with the $q_6$ order parameter, we observe that structural changes occur independently in the two domains, with the minor domain always more likely to fluctuate than the major domain (Fig. 4e and Supplementary Note 5). Furthermore, the relative difference in order between the two domains increases as etching proceeds. Once 750 atoms have been etched from the initial nanocrystal, the barrier to collective rearrangements in the minor domain lowers, allowing the twin boundary to relax into a single crystal (Fig. 4f).

The changes in vibrational free energy calculated over the course of the MD trajectories confirm that the transitions both from twinned to disordered and from disordered to single crystalline are thermodynamically favorable and driven by the relaxation of strain originating from the grain boundary (Extended Data Fig. 8).

## Conclusion

In summary, we directly visualized atomic dynamics of Au nanocrystals in reactive liquid, revealing structural fluctuations between ordered and disordered arrangements during solution-phase etching. These findings reveal the intrinsic dynamical nature of colloidal nanocrystals in reactive liquids. The ability of our approach to extract meaningful structural information from noisy, extreme imaging conditions positions it as useful in *in situ* EM and other imaging techniques. Fields requiring direct visualization of atomic- and molecular-scale structural changes and millisecond dynamics, from energy and catalytic materials to soft-matter and biological systems, stand to benefit from this methodology.